\documentclass[twocolumn,english,pra,amssymb,aps,superscriptaddress,showpacs,twocolumn,amsmath,showkeys,floatfix,]{revtex4-1}
\usepackage[T1]{fontenc}
\usepackage[latin9]{inputenc}
\usepackage{geometry}
\usepackage[active]{srcltx}
\usepackage{amsmath}
\usepackage{graphicx}
\usepackage{esint}
\usepackage{epsfig}
\usepackage{epstopdf}
\usepackage{physics}
\usepackage{color}
\usepackage{natbib}
\usepackage{upgreek}
\makeatletter
\usepackage{babel}

\makeatother

\usepackage{babel}
\begin{document}

\title{Quantum properties of light propagating in a Coherent Population Oscillation Storage Medium}

%%%% Taille faisceau
%%%%% Gamma et Gamma_0

\author{P. Neveu}
\affiliation{Laboratoire Aim\'e Cotton, Universit\'e Paris-Sud, ENS Paris-Saclay,
CNRS, Universit\'e Paris-Saclay, 91405 Orsay, France}
\author{F. Bretenaker}
\affiliation{Laboratoire Aim\'e Cotton, Universit\'e Paris-Sud, ENS Paris-Saclay,
CNRS, Universit\'e Paris-Saclay, 91405 Orsay, France}
\affiliation{Light and Matter Physics Group, Raman Research Institute, Bangalore 560080, India}
\author{F. Goldfarb}
\affiliation{Laboratoire Aim\'e Cotton, Universit\'e Paris-Sud, ENS Paris-Saclay,
CNRS, Universit\'e Paris-Saclay, 91405 Orsay, France}
\author{E. Brion}
\affiliation{Laboratoire Aim\'e Cotton, Universit\'e Paris-Sud, ENS Paris-Saclay,
CNRS, Universit\'e Paris-Saclay, 91405 Orsay, France}
\affiliation{Laboratoire Collisions Agr\'egats R\'eactivit\'e, IRSAMC \& UMR5589 du CNRS, Universit\'e de Toulouse III Paul Sabatier, F-31062 Toulouse, France}

\begin{abstract}
We study the propagation and storage of a quantum field using ultra-narrow Coherent Population Oscillations (CPO) in a $\Lambda-$type atomic medium. The predictions for classical fields are checked experimentally in a metastable vapor at room temperature. We derive the evolution of its squeezing spectrum in the presence of a large classical pump field which enables CPO to exist. We show that the spontaneous emission of the residual population pumped into the excited state progressively destroys the quantum noise properties of the quantum field along propagation. The output quantum field therefore tends to be a coherent state, discarding the possibility to store quantum states of light with CPO.
\end{abstract}

\pacs{42.50.Lc, 42.50.Gy, 42.50.Ct}

\maketitle

\section{Introduction}

One of the prerequisites for implementing quantum information processing is the availability of quantum memories, i.e. quantum devices able to faithfully store quantum states and release them on demand with high fidelity \cite{Lvovsky09}. Since photons appear as natural information carriers, much efforts have been devoted to the development of quantum memories for light states during the last twenty years. The most common protocol exploits the strong dispersion which arises together with the Electromagnetically Induced Transparency (EIT) phenomenon: a very narrow transmission resonance can be obtained when two optical transitions couple two ground states to the same excited state in a $\Lambda-$system \cite{Harris90, Lukin00}. Using this two-photon resonance, light pulses can be stored in the Raman coherence between both lower states of such a $\Lambda-$system in cold atoms \cite{Liu01} or atomic vapors \cite{Phillips01,Reim11} using close to or far off optical resonance schemes, as well as in ion Coulomb crystals \cite{Magnus11} or rare earth ions in matrices \cite{Longdell05}. EIT-based storage in warm vapors was also demonstrated to preserve single photon \cite{Eisaman05} or squeezed \cite{Appel08} states of light. Propagation under such EIT conditions can be described as the interplay between a light field and the Raman coherence, embodied by the dark state polariton \cite{Fleischhauer00}. One drawback of this storage protocol is the very high sensitivity of the coherence, and thus of the dark state polariton, to dephasing effects induced by the environment, which quickly destroy the memory.

Recently, Coherent Population Oscillations (CPO) have been shown to offer an alternative way to efficiently store classical light pulses in a $\Lambda$-system \cite{Eilam10,Maynard14,Almeida14}. Their phase preservation properties even allowed to store and retrieve orbital angular momentum of light \cite{Almeida15}. The physics of the CPO phenomenon is very different from EIT. It was first identified in a two-level system excited close to optical resonance by two coherent light fields, a strong one called the pump and a weaker one called the probe  \cite{Boyd81}. When these fields are slightly detuned from each other, intensity beats are induced. If this intensity modulation is slow enough, i.e. if the beat frequency is smaller than the excited state decay rate, atomic populations then experience a dynamical saturation and adiabatically follow the intensity variations. This population difference modulation leads to an amplification of the light modulation depth. This effect can be seen as a transmission window being opened for the probe beam within the absorption linewidth of the transition. CPO resonances have been observed in particular in solid-state systems \cite{Lee80,Chang04,Mrozek16,AY09}, with linewidths limited by the upper level decay rate: they are thus usually larger than EIT resonances between ground or metastable states, which do not experience spontaneous emission. However, the CPO resonance can become dramatically narrow in a $\Lambda$ system, when two antiphase CPO phenomena occur along the two legs and combine in an effective so-called ultra-narrow CPO between the two long lived ground states of the system \cite{Laupretre12}. As in the EIT case, it is possible to model the propagation under such CPO conditions as the interplay between a light field and the ground states population imbalance \cite{Neveu17} embodied by the so-called populariton. As populations are not sensitive to dephasing effects contrary to Raman coherences, CPO-based storage of classical light fields was demonstrated to be quite immune to perturbations such as magnetic field inhomogeneities \cite{Maynard14}. Nevertheless, the question whether quantum properties can be preserved or not in a CPO-based storage protocol has not been addressed yet. The aim of the present paper is thus to investigate the noise properties of a quantum light field propagating under such ultra-narrow CPO conditions.

To answer this question, Sec.\,\ref{sectionII} presents the $\Lambda$-system and the excitation scheme, together with the experimental results obtained in metastable helium for a weak classical probe field. These observations are in excellent agreement with the theory published in \cite{Neveu17}. Section\,\ref{sectionIII} extends the previous semi-classical theoretical treatment to quantized states of  probe field light, while the driving field remains classical. We then derive the evolution of the probe field quantum noise along propagation, with methods similar to the ones used in \cite{Peng05}. This approach is then applied in Sec.\,\ref{sectionIV} to a first-order derivation of the modifications of the field fluctuations due to the interaction with the medium under ultra-narrow CPO conditions. In particular, we investigate the role of the small population remaining in the excited state, which induces a detrimental additive noise through spontaneous emission.

\section{Classical behavior}
\label{sectionII}
\label{PartExp}
In this section, we first experimentally test the classical model derived in \cite{Neveu17}. A $\Lambda$-system composed of two $\sigma^+$ and $\sigma^-$ transitions is excited by an electric field propagating along the $z$-direction given by: 
\begin{equation}
\mathbf{E}\left(z,t\right)=\frac{\hbar}{d}\left(\Omega_D\left(z,t\right)\mathbf{e}_{||}+g\mathcal{E}\left(z,t\right)\mathbf{e}_\perp\right)e^{-\mathrm i \omega_0 \left(t-\frac{z}{c}\right)}+\mathrm{h.c}\,,
\label{TotalField}
\end{equation}
where $\Omega_D$ is the Rabi frequency of the monochromatic driving pump field at frequency $\omega_0$ and $\mathcal{E}$ the dimensionless envelope of the weaker field that we want to store. The two fields can oscillate at two different optical frequencies, since $\mathcal E$ can be time-dependent in a frame rotating at $\omega_0$ (see Appendix \ref{appendixA} for the details of the notations). The quantity $g=d\sqrt{\omega_{0}/2\hbar\epsilon_{0}V}$ holds for the atom-light dipolar coupling strength, where $d$ is the transition dipole moment and $V$ the field quantization volume.

In such a system, the transmission of a classical input probe field depends on its relative phase $\Theta$ with respect to the pump field. If the probe spectrum is symmetric with respect to the pump frequency $\omega_0$ and fits within the CPO linewidth, the phase sensitive transmission coefficients $T_{\Theta=0}$ and $T_{\Theta=\frac{\pi}{2}}$ are given by \cite{Neveu17}:
\begin{align}
T_{\Theta=\frac{\pi}{2}} & =\exp\left[\frac{g^2N}{2\Gamma c}\int_{0}^{L}\mathrm{d}z\left(\frac{2s\left(z\right)}{\frac{\gamma_{t}}{\Gamma_{0}}+3s\left(z\right)}-1\right)\right.\nonumber\\
&\qquad\qquad\qquad\qquad\qquad\left.\times\frac{1}{1+3s\left(z\right)}\right]\,,\\
T_{\Theta=0} & =\exp\left[-\frac{g^2N}{2\Gamma c}\int_{0}^{L}\frac{\mathrm{d}z}{1+3s\left(z\right)}\right]\,,
\label{EqPQT}
\end{align}
%\begin{align}
%T_{\Theta=\frac{\pi}{2}} & =\exp\left[\frac{g^2N}{2\Gamma %c}\int_{0}^{L}\mathrm{d}z\left(\frac{2s\left(z\right)}{3\frac{\gamma_{t%}}{\Gamma_{0}}+s\left(z\right)}-1\right)\frac{1}{1+s\left(z\right)}\right]\,,\\
%T_{\Theta=0} & =\exp\left[-\frac{g^2N}{2\Gamma %c}\int_{0}^{L}\frac{\mathrm{d}z}{1+s\left(z\right)}\right]\,,
%\label{EqPQT}
%\end{align}
where $N$ is the number of atoms interacting with the field, $\gamma_t$  the transit-induced decay and feeding rate of the lower levels population, $\Gamma_0$  the spontaneous emission decay rate of the upper level, $\Gamma$  the optical coherence decay rate and $s(z)=\Omega_D^2\left(z\right)/\Gamma\Gamma_0$  the saturation parameter of the transitions (see Appendix \ref{appendixB} for more details). Although $T_{\Theta=0}$ is the usual nonlinear absorption of a saturated transition, $T_{\Theta=\pi/2}$ has a more complex shape, because of the ultranarrow CPO contribution.

\begin{figure}
\epsfig{file=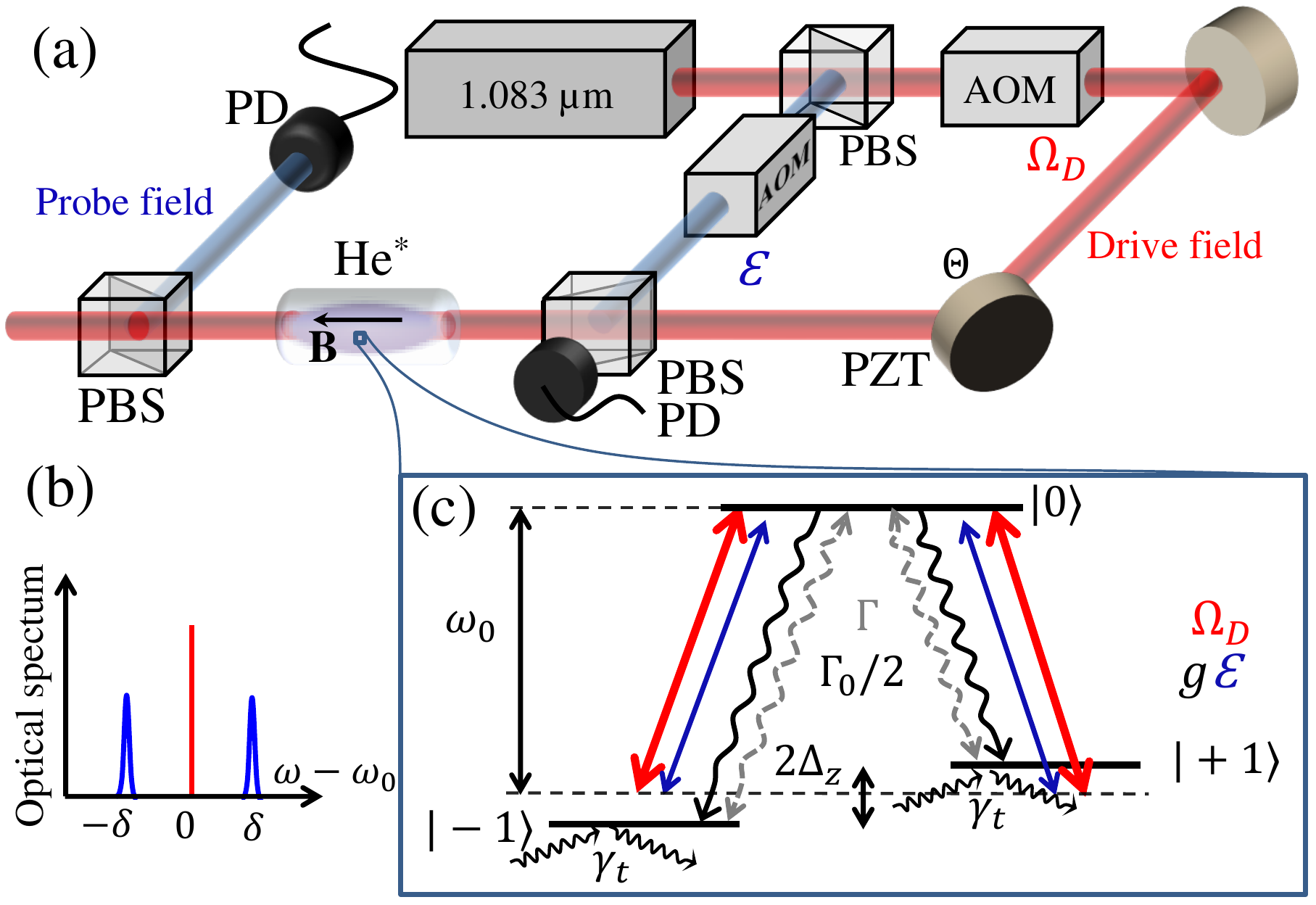,width=\columnwidth}
\caption{(a) Experimental setup. The laser is split by a polarizing beamsplitter (PBS) to obtain the orthogonally polarized pump ($\Omega_D$, red) and probe ($\mathcal{E}$, blue) fields. Their temporal and spectral shapes are controlled using acousto-optic modulators (AOMs): the probe spectrum (b) has two components symmetrically detuned by $\pm\delta$ from the coupling frequency. The relative phase $\Theta$ between the probe and pump fields is scanned with a mirror mounted on a piezoelectric actuator (PZT), and measured at the entrance of the cell. At the cell output, a third PBS isolates the probe from the driving field before detection. (c) Level structure. The $2^3S_1\leftrightarrow2^3P_1$ transition excited by linearly polarized light ends up in a closed $\Lambda$ structure \cite{Maynard14}. $\Gamma_0$, $\Gamma$ and $\gamma_t$ are the decay rates of the excited state population, the optical coherences, and the lower levels populations, respectively.\label{Setup}}
\end{figure}

To check this model, we measure the transmission of a weak classical probe field under such ultra-narrow CPO conditions, using the setup described in Fig.\,\ref{Setup}(a). Since we need the pump and probe fields to be coherent for the CPO process to occur, they are both derived from the same laser source and separated by a polarizing beamsplitter (PBS). Two different acousto-optic modulators (AOMs) allow to independently control their amplitudes and frequencies. Here we investigate the situation depicted in Fig.\,\ref{Setup}(b) where the probe spectrum consists in two tones, called signal and idler, respectively detuned by $+\delta$ and $-\delta$ with respect to the pump field. The relative phase $\Theta$ between the probe and pump fields is scanned thanks to a mirror mounted on a piezoelectric actuator (PZT) and placed in the path of the pump field. A second PBS recombines the fields at the entrance of the helium cell. A small part of the fields, exiting the other port of the PBS, is detected to monitor $\Theta$. After propagation in the 6-cm-long cell filled with 1\,Torr of helium, the probe field is isolated from the pump field by a third PBS before detection. The cell is protected from stray magnetic fields by a $\mu$-metal shield, and a longitudinal magnetic field is applied to lift the degeneracy between the Zeeman sub-levels. Inside the cell, the $1/e^2$ waists of the drive and probe beams are 2\,mm.

\begin{figure}
\epsfig{file=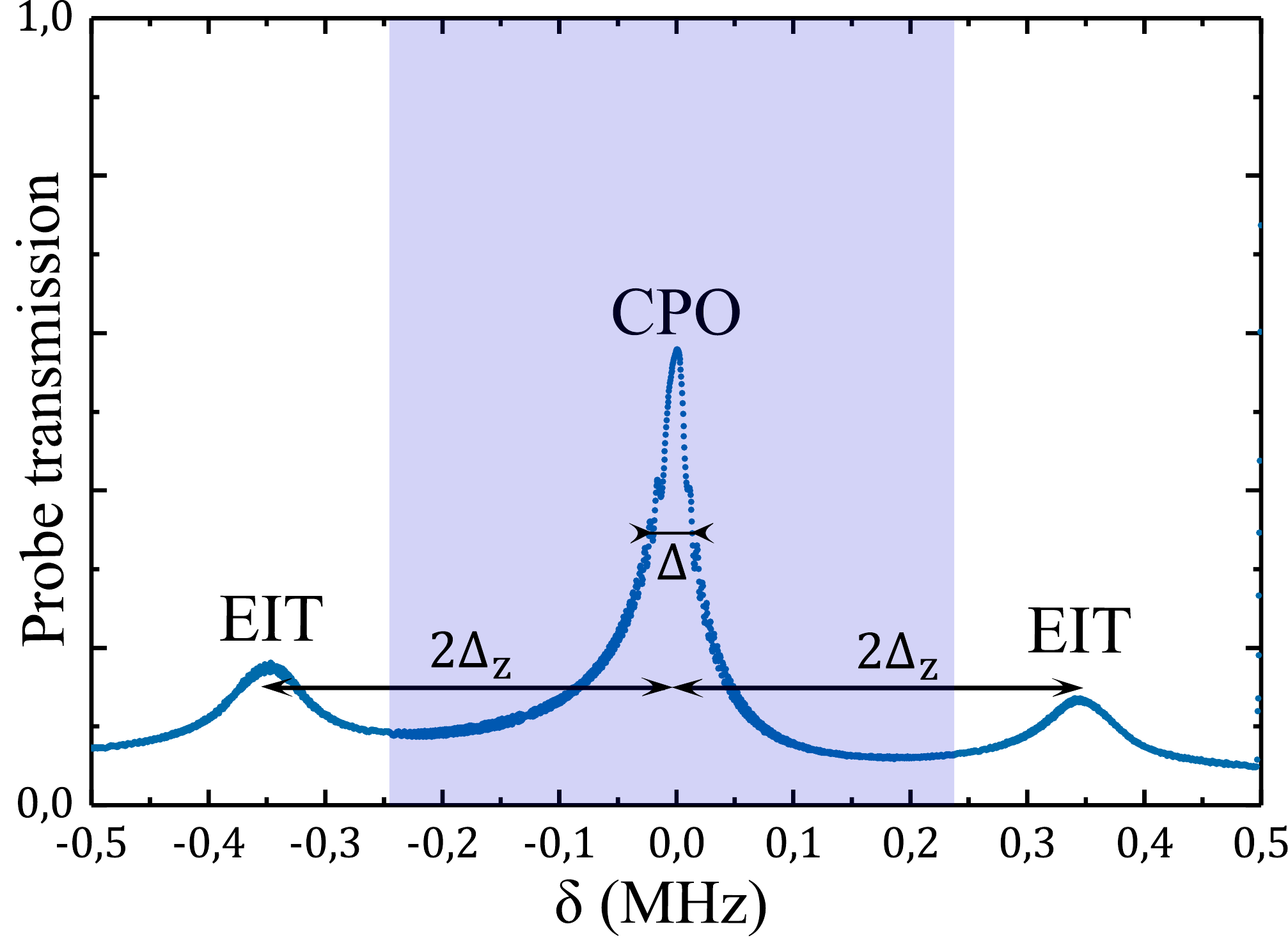,width=\columnwidth}
\caption{Measurement of a 100\,$\upmu$W probe field absorption spectrum under CPO conditions, in presence of a 10 mW driving field. Within the nearly 1\,GHz Doppler-broadened absorption width of the transition, three transmission resonances are visible. A small longitudinal magnetic field shifts the EIT resonances $\pm2\Delta_z$ away from the central CPO resonance. In our experimental conditions, the spectrum of the signal is fully included within the shadowed region, where no Raman coherence is excited. \label{FigSpectrum}}
\end{figure}

Figure \ref{Setup}(c) shows the excitation scheme of the $2^3S_1\leftrightarrow2^3P_1$ transition of metastable helium. The quantization axis of the atomic levels is chosen along the propagation direction of light. The excited and ground levels are composed of three Zeeman sub-levels. Since the transition $m=0\leftrightarrow m=0$ is forbidden when the levels have the same total momentum $J$, the system ends up in a closed $\Lambda$ scheme composed by the $m=\pm1$ ground states and the $m=0$ excited state \cite{Goldfarb08}. One has experimentally $\Gamma/2\pi\simeq 0.8\,$GHz, $\Gamma_0/2\pi=1.6\,$MHz and $\gamma_t/2\pi\simeq20\,$kHz.

The total field $\mathbf{E}$ can be decomposed in the circular basis $\{\mathbf{e}_+,\mathbf{e}_-\}$ using
\begin{equation}
\mathbf{E}\cdot\mathbf{e}_\pm=\frac{\mathbf{E}\cdot\mathbf{e}_{||}\pm \mathrm{i}\mathbf{E}\cdot\mathbf{e}_{\perp}}
{\sqrt 2},\label{PolarDecomposition}\end{equation}
where $\mathbf e_{||}$ and $\mathbf e_{\perp}$ are the crossed linear polarization directions of the pump and probe fields, respectively. The system thus experiences balanced excitations along the two legs of the $\Lambda$-system: the $m=-1\leftrightarrow m=0$ (resp. $m=1\leftrightarrow m=0$) transition is excited by a beatnote due to the $\sigma^+$ (resp. $\sigma^-$) components of the probe and  pump fields, so that CPOs occur between the $\ket{-1}$ (resp. $\ket{1}$) ground state and the $\ket{0}$ excited state. Depending on the relative polarization angle and on the relative phase $\Theta$ between the two fields, symmetric or antisymmetric combinations of CPOs between both legs of the system can be excited. In particular, when the two fields are orthogonally polarized, a phase difference $\Theta=\pi/2\,\left[\pi\right]$ excites the antisymmetric mode, in which the two CPO phenomena are in antiphase. The population then oscillates between the two ground states of the system. This also leads to an ultranarrow transparency window for the two-frequency probe, centered on the pump field frequency (see Fig.\,\ref{FigSpectrum}). On the contrary, when $\Theta=0$, only probe absorption remains.

For degenerate Zeeman ground states, the pump (resp. probe) field exciting the left leg of the $\Lambda$ system and the probe (resp. pump) field along the right leg lead to a two-photon EIT resonance when both light fields have the same frequency. This  also corresponds to the situation where the CPO resonance condition is fulfilled. In order to get rid of EIT two-photon resonance, we apply a longitudinal magnetic field to the atoms. Then, the Zeeman shift $2\Delta_{z}$ between the $m=\pm1$ ground-state sub-levels restrains Raman coherence from being excited, provided the probe spectrum fits within a window of width smaller than $4\Delta_z-W_\textrm{EIT}$, where $W_\textrm{EIT}$ is the EIT linewidth. Figure\;\ref{FigSpectrum} shows an experimental transmission spectrum of a single-frequency probe field in such conditions: in the center of the nearly 1\,GHz Doppler broadened absorption window, one can see a CPO resonance surrounded by two EIT resonances, shifted because of the Zeeman shift. When the probe spectrum is fully confined to the shadowed region of this plot, EIT is avoided and only ultra-narrow CPO occurs. 

\begin{figure}
\epsfig{file=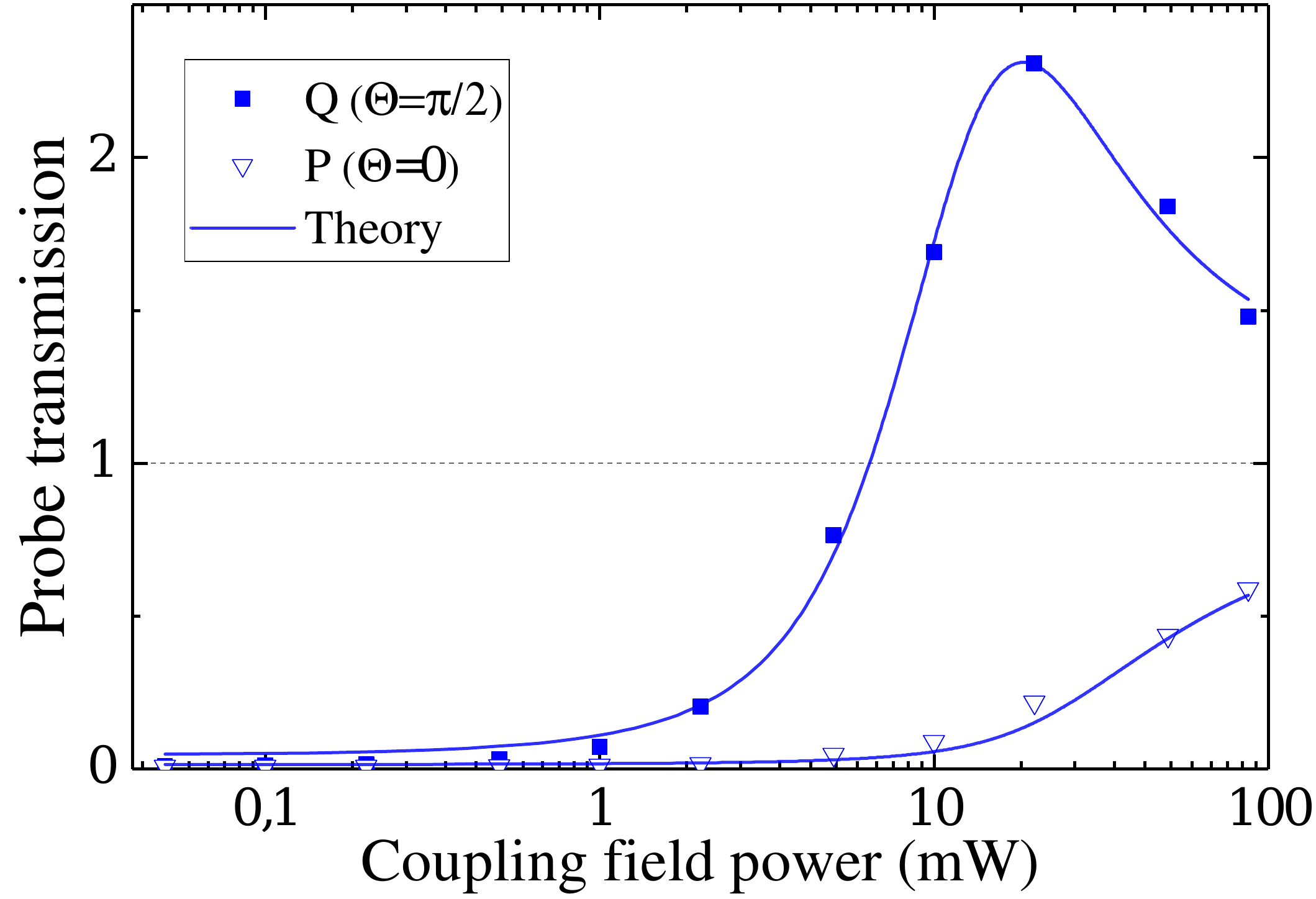,width=\columnwidth}
\caption{Measured evolution of the evolution of the probe field transmission versus pump field power. Filled squares (resp. open triangles): $\Theta=\pi/2$ (resp. $\Theta=0$ s). The input probe field contains two spectral components at $\nu=\pm2\,$kHz, with equal amplitudes. Full line: theory based on Eqs.\,\ref{EqPQT} taking into account the $\sim20\,\%$ residual absorption from the $2^3S_1\leftrightarrow2^3P_2$ transition.\label{FigPQT}}
\end{figure}

Figure\,\ref{FigPQT} shows a comparison between the theoretical transmission coefficients for the two field quadratures quadratures $P$ and $Q$ obtained from Eq.\,\ref{EqPQT} with $\Theta=0$ and $\pi/2$ respectively and  the experimental data. An excellent agreement is observed. Below an input optical pump field power of $\sim$3\,mW, ultra-narrow CPO cannot be excited because the saturation induced by the pump field is too weak. Above an input optical pump power of $\sim$30\,mW, the strong saturation of the atoms by the pump makes the medium transparent for the probe field and prevents the CPO resonance from appearing. In between these two regimes, CPO is efficiently excited and a strong phase sensitive behavior takes place. The values of the fitting parameters used to plot the continuous lines of Fig.\,\ref{FigPQT} are $\gamma_t/\Gamma_0=9.6(9)\times10^{-2}$, $g^2NL/2\Gamma c=2.8(1)$, and $s/P_\mathrm {opt}=4.7(5)\times10^{-1}\,\mathrm{W^{-1}}$. Finally, a residual absorption of $20(2)\,\%$ induced by the by the $D_2$ transition, which is detuned by 2.3\,GHz, is taken into account.

\section{Quantum treatment of the probe propagation}
\label{sectionIII}
Let us now study the evolution of the quantum properties of the probe field along propagation in the presence of a steady-state pump. To this aim, we consider the spectral complex Fourier amplitudes of the probe field operator $\mathcal E\left(z,t\right)$ and of its Hermitian conjugate $\mathcal E\left(z,t\right)^\dagger$:
\begin{align}
{\cal E}\left(z,\nu\right) &=\frac{1}{\sqrt{2\pi}}\int_{-\infty}^{+\infty}{\cal E}\left(z,t\right)e^{\mathrm{i}\nu t}\mathrm{d}t,
\label{DefE}
\end{align}
\begin{align}
{\cal E}^{\dagger}\left(z,\nu\right)&=\frac{1}{\sqrt{2\pi}}\int_{-\infty}^{+\infty}{\cal E}\left(z,t\right)^{\dagger}e^{\mathrm{i}\nu t}\mathrm{d}t.
\label{DefEdagger}
\end{align}
It is important to notice that $\mathcal E\left(z,\nu\right)$ and $\mathcal E^\dagger\left(z,\nu\right)$ are not Hermitian conjugate of each other.

The equations of evolution are derived in the Heisenberg picture but, to avoid clumsy notations, the time dependence is not explicitly written. The quantum fluctuations originating from the coupling of the atoms to the vacuum electromagnetic bath result in Langevin forces, which add to the Heisenberg time evolution equations for the operators \cite{Cohen}. We assume that spontaneous emission is the only source of decoherence, so that $\Gamma=\Gamma_{0}/2$.

\subsection{Evolution of the atoms}
We consider an ensemble of atoms described by a $\Lambda$-scheme similar to the one of Fig.\,\ref{Setup}(c). Two ground states $\left|-1\right\rangle$ and $\left|+1\right\rangle$ are optically coupled to the same excited state $\left|e\right\rangle$, both transitions having the same frequency $\omega_{0}$. The atom $j$ is described by the set of operators $\sigma^j_{\mu\nu}$, defined in the frame rotating at $\omega_0$ as:
\begin{equation}
\sigma^j_{\nu\nu}=\left|\nu\right\rangle _{jj}\left\langle \nu\right| \textrm{ and } \sigma^j_{\pm1\mp1}=\left|\pm1\right\rangle _{jj}\left\langle \mp1\right|\ ,
\end{equation}
where $\nu\in\{e,-1,+1\}$ and, for the optical coherences:
\begin{equation}
\sigma_{\pm1e}^{j}=\left|\pm1\right\rangle _{jj}\left\langle e\right|e^{\mathrm{i}\omega_{0}\left(t-z/c\right)}.
\end{equation}

When a longitudinal magnetic field is applied along the $z$ axis and when the pump and probe fields propagate into the medium, the Hamiltonian in the rotating wave approximation is
$\mathrm{H}^j=\mathrm{H}^j_{\cal Z}+\mathrm H^j_d$, where
\begin{equation}
\mathrm{H}^j_{\cal Z}=\hbar\Delta_{z}\left(\sigma^j_{11}-\sigma^j_{-1-1}\right)
\end{equation}
corresponds to the Zeeman interaction with the magnetic field, shifting the ground states by $\pm\Delta_z$, and
\begin{equation}
\mathrm H^j_d=\hbar\left(\sigma^j_{e1}\mathbf{e_{-}}+\sigma^j_{e-1}\mathbf{e_{+}}\right)\left(\Omega_D\mathbf{e_{||}}+g{\cal E}\mathbf{e_{\perp}}\right)+\mathrm{h.c.}
\end{equation}
is the electric-dipole interaction with the optical fields.

Rather than considering individual atomic operators, we assume that the medium is homogeneous and define continuous $z$-dependent operators $\sigma_{\mu\nu}\left(z\right)$ by averaging the density operator components over a thin slice of medium $\mathcal{T}\left(z\right)$ containing $N$ atoms:
\[
\sigma_{\mu\nu}\left(z\right)=\frac{1}{N}\sum_{j\in\mathcal{T}\left(z\right)}\sigma_{\mu\nu}^{j}.
\]
This approximation is valid when the width of the slice
is large enough to contain a large number of atoms, but small enough compared to the light wavelength, so that it is possible to differentiate on $z$. The Heisenberg-Langevin equations, which govern the dynamics of the atomic operators $\sigma_{\mu\nu}$, are then \cite{Cohen}
\begin{equation}
\frac{\partial}{\partial t}\sigma_{\mu\nu}=\frac{1}{\mathrm{i}\hbar}\left[\sigma_{\mu\nu},\mathrm{H}\right]+{\cal R}\left(\sigma_{\mu\nu}\right)+F_{\mu\nu},
\label{HL}
\end{equation}
where $\mathrm{H}$ is the sum of the $\mathrm{H}^j$'s in the slice $\mathcal T\left(z\right)$, $\mathcal{R}$ is the spontaneous emission dissipator and $F_{\mu\nu}$ is the Langevin noise operator, spatially averaged in $\mathcal T\left(z\right)$:
\begin{equation}
F_{\mu\nu}\left(z,t\right)=\frac{1}{N}\underset{j\in{\cal T}\left(z\right)}{\sum}F_{\mu\nu}^{j}\left(t\right).
\end{equation}

The average of the Langevin forces is zero, and we assume that their correlation timescale can be neglected with respect to the timescales of the dynamics of the system, so that:
\begin{equation}
\left\langle F_{\mu\nu}\left(z_1,t_1\right)F_{\alpha\beta}\left(z_2,t_2\right)\right\rangle=\frac{\delta_{z_1}^{z_2}}{N}D_{\mu\nu}^{\alpha\beta}\delta\left(t_1-t_2\right),
\label{Langevin}
\end{equation}
where $\delta_{z_1}^{z_2}$ is a Kronecker symbol equal to zero for two different spatial positions, and $D_{\mu\nu}^{\alpha\beta}$ is a diffusion coefficient, given by the Einstein generalized relations \cite{Cohen}
\begin{equation}
D_{\mu\nu}^{\alpha\beta}=\left\langle\mathcal{R}\left(\sigma_{\mu\nu}\sigma_{\alpha\beta}\right)-\sigma_{\mu\nu}\mathcal{R}\left(\sigma_{\alpha\beta}\right)-\sigma_{\alpha\beta}\mathcal{R}\left(\sigma_{\mu\nu}\right)\right\rangle.
\label{Einstein}
\end{equation}

\subsection{Evolution of the optical fields}

The total optical field $\mathbf E$ propagates inside the medium according to Maxwell's equations in the slowly varying envelope approximation. It is possible to derive separate equations of for the pump and probe field envelopes by projecting the equation for the total field on their respective orthogonal polarization directions using Eqs.\,(\ref{TotalField}) and (\ref{PolarDecomposition}), leading to:
\begin{align}
\left(c\partial_z+\partial_t\right)\Omega_D&=\mathrm ig^2N\left(\sigma_{e1}+\sigma_{e-1}\right)\ , \nonumber\\
\left(c\partial_z+\partial_t\right)\mathcal{E}&= gN\left(\sigma_{e1}-\sigma_{e-1}\right)\ .
\label{MW}
\end{align}

As shown in \cite{Neveu17}, the quadrature operators of the complex amplitude $\mathcal E$ defined by
\begin{align}
{\cal P}\left(z,t\right)=\frac{1}{2}\left({\cal E}\left(z,t\right)+{\cal E}\left(z,t\right)^{\dagger}\right) \nonumber\\
{\cal Q}\left(z,t\right)=\frac{1}{2\mathrm{i}}\left({\cal E}\left(z,t\right)-{\cal E}\left(z,t\right)^{\dagger}\right)
\end{align}
are the relevant quantities to describe the probe field. They coincide with the usual probe quadratures as soon as the probe spectrum is symmetric and centered on $\omega_0$. For instance, $\cal P$ and $\cal Q$ are the usual quadratures for a monochromatic field at $\omega_0$. It is also the case for a probe containing monochromatic signal and idler components of equal amplitudes located at frequencies $\omega_0\pm\delta$, respectively.

\subsection{Linearization}

The preceding equations can be solved using a perturbative approach at first order in quantum probe field. Any observable $\cal O$ can then be expanded as follow:
\begin{equation}
\mathcal O=\left\langle \mathcal O\right\rangle _{0}+\left\langle \mathcal O\right\rangle _{1}+\delta\mathcal O
\end{equation}
where $\left\langle \mathcal O\right\rangle _{0}$ stands for the mean value of $\mathcal O$ in the presence of the classical pump field alone, $\left\langle \mathcal O\right\rangle _{1}$ is the first order perturbation due to the presence of the probe field and $\delta \mathcal O$ represents the linearized quantum fluctuation part of $\mathcal O$.

Notice that we neglect the influence of the probe field on the evolution of the pump field, and we assume that the evolution of the quantum fluctuations of any operator is governed by the dynamics generated by the pump field only.

\section{Results}
\label{sectionIV}
\subsection{Zeroth order}

The zeroth-order dynamics is obtained from the mean values of Eqs.\,(\ref{HL},\ref{MW}) in the presence of the pump alone, reading
\begin{align}
0&=\frac{1}{\mathrm{i}\hbar}\left[\left\langle \sigma_{\mu\nu}\right\rangle _{0},\left\langle\mathrm H\right\rangle_0\right]+{\cal R}\left(\left\langle \sigma_{\mu\nu}\right\rangle _{0}\right)\ , \nonumber\\
c\partial_z\Omega_D&=\mathrm ig^2N\left(\left\langle \sigma_{e1}\right\rangle_0 +\left\langle \sigma_{e-1}\right\rangle_0 \right)\ ,
\end{align}
where the time dependence is skipped because the pump is assumed to be steady.

Assuming that the pump Rabi frequency is real, the density matrix of the system is then given by
\begin{equation}
\left\langle \sigma\right\rangle _{0}=\left(\begin{array}{ccc}
\frac{s}{1+3s} & \frac{\mathrm{i}\Omega_{D}}{\Gamma_{0}\left(1+3s\right)} & \frac{\mathrm{i}\sqrt{2}\Omega_{D}}{\Gamma_{0}\left(1+3s\right)}\\
-\frac{\mathrm{i}\sqrt{2}\Omega_{D}}{\Gamma_{0}\left(1+3s\right)} & \frac{1+2s}{2+6s} & 0\\
-\frac{\mathrm{i}\sqrt{2}\Omega_{D}}{\Gamma_{0}\left(1+3s\right)} & 0 & \frac{1+2s}{2+6s}
\end{array}\right)\ .
\end{equation}
Such a density matrix merely describes the usual saturation of the transition, leading to a nonzero amount of population in the excited state. The absorption of the pump leads to a $z-$dependent $s$ parameter, which obeys the following equation:
\begin{equation}
\partial_{z}s=-\frac{2g^{2}N}{c\Gamma_{0}}\ \frac{s}{1+3s}\ .
\label{PropaS}
\end{equation}

\subsection{First order -- Expectation value}

Eqs.\,(\ref{HL},\ref{MW}) in the Fourier domain give the following set of equations for the probe field expectation value:
\begin{align}
\begin{split}
-\mathrm i \nu\partial_t\left\langle\sigma_{\mu\nu}\right\rangle_1=&\frac{1}{\mathrm{i}\hbar}\left(\left[\left\langle \sigma_{\mu\nu}\right\rangle _{1},\left\langle\mathrm H\right\rangle_0\right]+\left[\left\langle \sigma_{\mu\nu}\right\rangle _{0},\left\langle\mathrm H\right\rangle_1\right]\right)\ , \nonumber\\
&+{\cal R}\left(\left\langle \sigma_{\mu\nu}\right\rangle _{1}\right)
\end{split}\\
\left(c\partial_z-\mathrm i \nu\right)\left\langle\mathcal E\right\rangle=&\mathrm gN\left(\left\langle \sigma_{e1}\right\rangle_1 +\left\langle \sigma_{e-1}\right\rangle_1 \right)\ .
\end{align}

When the probe field spectrum fits within the $4\Delta_z-W_\textrm{EIT}$ window centered on $\omega_0$, as shown in the Fig.\,\ref{FigSpectrum}, (assuming $\nu\ll\Delta_z\ll\Gamma$), the probe field complex amplitude propagates according to
\begin{equation}
\left(c\partial_{z}-\mathrm{i}\nu\right)\left\langle {\cal E}\right\rangle  = \frac{g^{2}N}{\Gamma_{0}\left(1+3s\right)}\frac{\mathrm{i}\nu\left\langle{\cal E}\right\rangle-\Delta\left\langle{\cal E}^{\dagger}\right\rangle}{\Delta-\mathrm{i}\nu}\ .
\end{equation}
Such an equation can be re-written using the Fourier components of the quadratures ${\cal P}\left(z,\nu\right)$ and ${\cal Q}\left(z,\nu\right)$:
\begin{align}
\partial_{z}\left\langle {\cal Q}\right\rangle  & = \left(\Lambda_{1}\left(z,\nu\right)+\mathrm i\frac{\nu}{c}\right)\left\langle {\cal Q}\right\rangle\ ,
\nonumber\\
\partial_{z}\left\langle {\cal P}\right\rangle  & = \left(\Lambda_{2}\left(z,\nu\right)+\mathrm i\frac{\nu}{c}\right)\left\langle {\cal P}\right\rangle\ ,
\label{PropaQP}
\end{align}
with $\Lambda_{1,2}$ given by
\begin{align}
\Lambda_{1}= &+ \frac{g^{2}N}{c\Gamma_{0}\left(1+3s\right)}\ \frac{\Delta+\mathrm{i}\nu}{\Delta-\mathrm{i}\nu}\ , \nonumber\\
\Lambda_{2}= & -\frac{g^{2}N}{c\Gamma_{0}\left(1+3s\right)}\ .
\label{Lambda}
\end{align}

Equations (\ref{PropaQP}) do not mix ${\cal P}\left(z,\nu\right)$ and ${\cal Q}\left(z,\nu\right)$, which are thus eigenmodes for the propagation. An adiabatic expansion of $\Lambda_1$ then gives back the classical CPO dispersion behavior:
\begin{equation}
\begin{split}
\Lambda_{1}+\mathrm i \frac{\nu}{c}\simeq & \frac{g^{2}N}{c\Gamma_{0}\left(1+3s\right)}+\mathrm{i}\nu\frac{1+\frac{g^{2}N}{\Omega_{D}^{2}\left(1+3s\right)}}{c}\\
&-\frac{g^{2}N}{c\Gamma_{0}\left(1+3s\right)}\frac{\nu^{2}}{\Delta^{2}}\ .\label{eq:DvptLambda}
\end{split}
\end{equation}
We can see that CPOs generate a phase sensitive behavior as soon as the spectrum of the probe fits within the CPO linewidth $\nu\ll\Delta$: the first term of Eq.\,(\ref{eq:DvptLambda}) corresponds to the amplification, the second one corresponds to the associated decrease in group velocity, and the third term limits the bandwidth for which these effects are efficient.

We now suppose that the probe spectrum is well within the CPO linewidth $\Delta$, with $\nu\ll\Delta$. Then, integrating Eq.\,(\ref{PropaQP}) and using Eqs.\,(\ref{PropaS},\ref{Lambda}), one can find the expectation values of the quadratures after propagation:
\begin{align}
\left\langle {\cal Q}\left(z,\nu\right)\right\rangle =&\sqrt{G\left(z\right)}e^{\frac{\mathrm{i}\nu z}{c}}\left\langle {\cal Q}\left(0,\nu\right)\right\rangle\ ,
\nonumber \\
\left\langle {\cal P}\left(z,\nu\right)\right\rangle =&\frac{1}{\sqrt{G\left(z\right)}}e^{^{\mathrm{i}\frac{\nu z}{c}}}\left\langle {\cal P}\left(0,\nu\right)\right\rangle\ ,
\end{align}
where $1/G\left(z\right)=s\left(z\right)/s\left(0\right)<1$ corresponds to the decrease of pump intensity due to absorption.

\subsection{First order -- Fluctuations}

The linearized equations of evolution for the fluctuation of the operators can be deduced from Eqs.\,(\ref{HL},\ref{MW}) in the Fourier domain: 
\begin{align}
\begin{split}
-\mathrm i \nu\partial_t\delta\sigma_{\mu\nu}=&\frac{1}{\mathrm{i}\hbar}\left(\left[\delta \sigma_{\mu\nu},\left\langle\mathrm H\right\rangle_0\right]+\left[\left\langle \sigma_{\mu\nu}\right\rangle _{0},\delta\mathrm H\right]\right)\ ,\\
&+{\cal R}\left(\delta \sigma_{\mu\nu}\right)+F_{\mu\nu}
\end{split}\\
\left(c\partial_z-\mathrm i \nu\right)\delta\mathcal E=&\mathrm gN\left(\delta\sigma_{e1}+\delta\sigma_{e-1}\right).
\end{align}
These equations are valid at first order and describe the effect of the pump saturation on the probe fluctuations. They neglect any effect of the probe evolution itself on its own fluctuations. The quadrature fluctuations are then given by
\begin{align}
\partial_{z}\delta{\cal Q} = & \left(\Lambda_{1}+\mathrm i\frac{\nu}{c}\right)\delta{\cal Q}+\underset{\mu\nu}{\sum}\alpha_{\mu\nu}F_{\mu\nu}
\ ,\nonumber\\
\partial_{z}\delta{\cal P} = & \left(\Lambda_{2}+\mathrm i\frac{\nu}{c}\right)\delta{\cal P}+\underset{\mu\nu}{\sum}\beta_{\mu\nu}F_{\mu\nu}\ ,
\label{PropaFluct}
\end{align}
with
\begin{equation}
\begin{array}{rcl}
\underset{\mu\nu}{\sum}\alpha_{\mu\nu}F_{\mu\nu}&= \frac{-gN}{\sqrt{2}c\Gamma_{0}\Delta}&\left[\nu\left(F_{e-1}-F_{e1}-F_{-1e}+F_{1e}\right)\right.
\\ & &\left.+\sqrt{2}\Omega_{D}\left(F_{11}-F_{-1-1}\right)\right]\ ,
\\
\underset{\mu\nu}{\sum}\beta_{\mu\nu}F_{\mu\nu}&= \frac{+gN}{\sqrt{2}c\Gamma_{0}}&\left(F_{e-1}-F_{e1}+F_{-1e}-F_{1e}\right).
\end{array}
\end{equation}

Three independent combinations of Langevin operators are thus relevant: $F_\Delta=F_{11}-F_{-1-1}$ and $F_{\pm}=F_{e1}\pm F_{e-1}$. Using Eqs.\,(\ref{PropaQP},\ref{PropaFluct}), it is possible to derive the quadratures after propagation:
\begin{align}
\begin{split}
{\cal Q}\left(z,\nu\right) =& \sqrt{G\left(z\right)}e^{\frac{\mathrm{i}\nu z}{c}}{\cal Q}\left(0,\nu\right)\\&
+\underset{\mu\nu}{\sum}\int_{0}^{z}\mathrm{d}xe^{\int_{x}^{z}\left(\Lambda_{1}+\mathrm i\frac{\nu}{c}\right)\mathrm{d}\xi}\alpha_{\mu\nu}F_{\mu\nu}\ ,
\label{FullQ}
\end{split}\\
\begin{split}
{\cal P}\left(z,\nu\right) =& \frac{1}{\sqrt{G\left(z\right)}}e^{^{\mathrm{i}\frac{\nu z}{c}}}{\cal P}\left(0,\nu\right)\\&+\underset{\mu\nu}{\sum}\int_{0}^{z}\mathrm{d}xe^{\int_{x}^{z}\left(\Lambda_{2}+\mathrm i\frac{\nu}{c}\right)\mathrm{d}\xi}\beta_{\mu\nu}F_{\mu\nu}\ .
\label{FullP}
\end{split}
\end{align}

\begin{table}[]
    \centering
\begin{tabular}{|c|c|c|c|}
\hline 
$D$    &
$...\left.F_{\Delta}(z,\nu)\right\rangle$   &
$...\left.F_{+}(z,\nu)\right\rangle$ &
$...\left.F_{-}(z,\nu)\right\rangle$
\tabularnewline
\hline 
$\left\langle F_{\Delta}(z,\nu)\right....$ & $\Gamma_{0}\frac{s}{1+3s}$ & $0$ & 0 \tabularnewline
\hline 
$\left\langle F_{+}(z,\nu)\right....$ & $0$ & 0& $0$ \tabularnewline
\hline
$\left\langle F_{-}(z,\nu)\right....$ & $0$ & 0 & $\Gamma_0$ \tabularnewline
\hline
\end{tabular}
    \caption{Diffusion coefficients associated with the correlations of two Langevin operators (see Eq.\,\ref{Langevin}).}
    \label{Tab}
\end{table}

The quadrature fluctuations after propagation can then be evaluated by computing their squeezing spectra, defined as the Fourier transform of the autocorrelation function. For a quadrature $X$, this spectrum given by:
\begin{equation}
\begin{split}
{\cal S}_{X}\left(z,\nu\right)\equiv&\frac{4c}{L}\int\mathrm{d}te^{\mathrm{i}\nu t}\left\langle X\left(z,t\right)X\left(z,0\right)\right\rangle \\=&\frac{4c}{L}\int\mathrm{d}\nu'e^{-\frac{\mathrm{i}\left(\nu+\nu'\right)z}{c}}\left\langle X\left(z,\nu\right)X\left(z,\nu'\right)\right\rangle\ , \label{eq:Flux}
\end{split}
\end{equation}
where the factor $4c/L$ is the vacuum shot noise renormalization factor (see Appendix \ref{appendixA}).

Applying Eq.\,(\ref{eq:Flux}) to $\cal P$ and $\cal Q$ leads to:
\begin{equation}
\begin{array}{l}
\mathcal S_\mathcal P \left(z,\nu\right)=\\
\quad\frac{4c}{L}\int\mathrm{d}\nu'\Big(e^{\int_{0}^{z}\left(\Lambda_{2}\left(\nu\right)+\Lambda_{2}\left(\nu'\right)\right)\mathrm{d}\xi}\left\langle {\cal P}\left(0,\nu\right){\cal P}\left(0,\nu'\right)\right\rangle\ \\
\\
\qquad\qquad+\underset{abcd}{\sum}\int_{0}^{z}\int_{0}^{z}\mathrm{d}x\mathrm{d}x'e^{\int_{x}^{z}\Lambda_{2}\left(\nu\right)\mathrm{d}\xi+\int_{x'}^{z}\Lambda_{2}\left(\nu'\right)\mathrm{d}\xi}\\
\qquad\qquad\times\beta_{ab}\beta_{cd}\left\langle F_{ab}\left(x,\nu\right)F_{cd}\left(x',\nu'\right)\right\rangle \Big)\ ,
\end{array}
\end{equation}
\begin{equation}
\begin{array}{l}
\mathcal S_\mathcal Q \left(z,\nu\right)=\\
\quad\frac{4c}{L}\int\mathrm{d}\nu'\Big(e^{\int_{0}^{z}\left(\Lambda_{1}\left(\nu\right)+\Lambda_{1}\left(\nu'\right)\right)\mathrm{d}\xi}\left\langle {\cal Q}\left(0,\nu\right){\cal Q}\left(0,\nu'\right)\right\rangle \\
\\
\qquad\qquad+\underset{abcd}{\sum}\int_{0}^{z}\int_{0}^{z}\mathrm{d}x\mathrm{d}x'e^{\int_{x}^{z}\Lambda_{1}\left(\nu\right)\mathrm{d}\xi+\int_{x'}^{z}\Lambda_{1}\left(\nu'\right)\mathrm{d}\xi}\\
\qquad\qquad\times\alpha_{ab}\alpha_{cd}\left\langle F_{ab}\left(x,\nu\right)F_{cd}\left(x',\nu'\right)\right\rangle \Big)\ .
\end{array}
\end{equation}

\begin{figure}
\epsfig{file=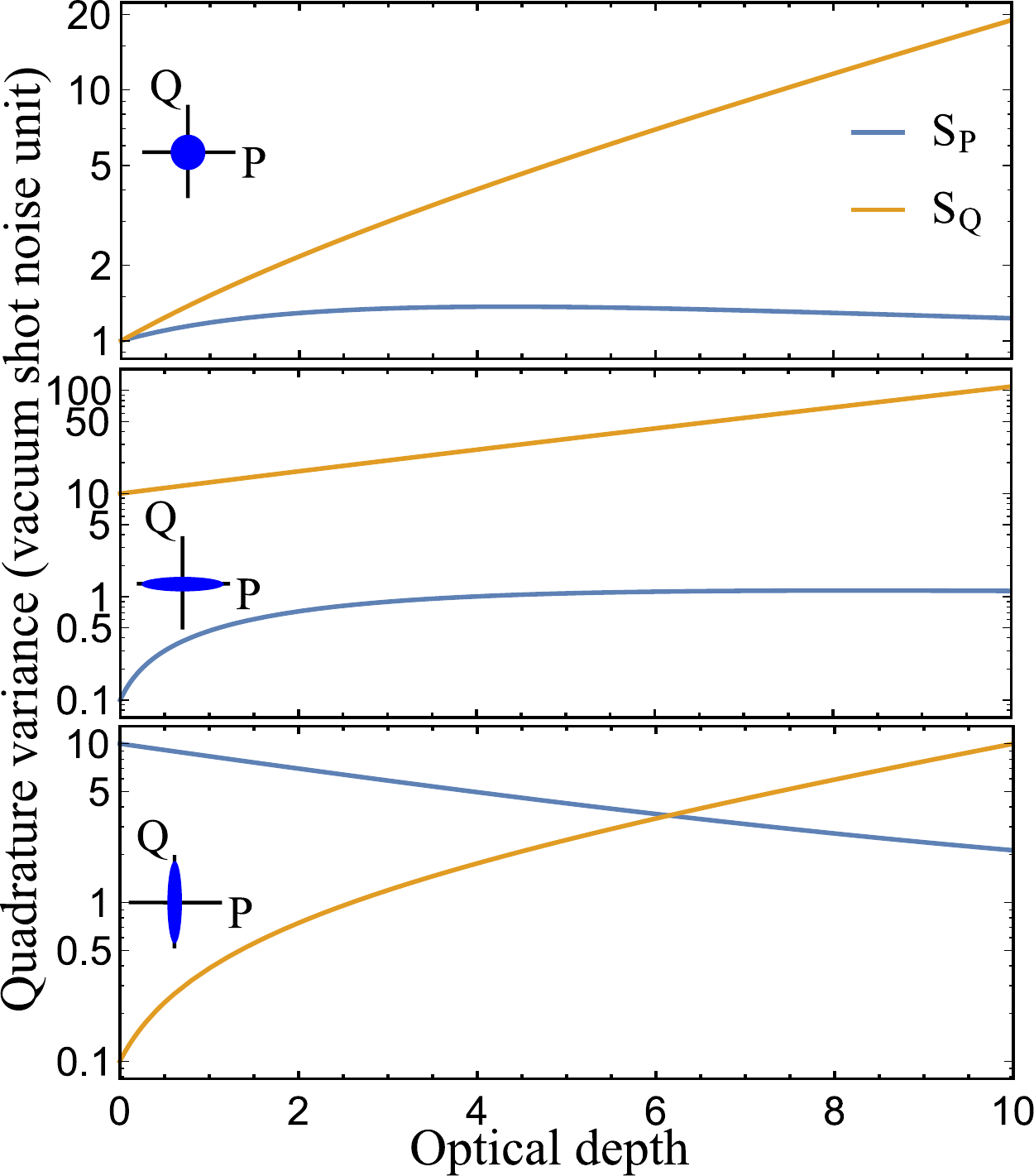,width=\columnwidth}
\caption{Evolutions of the variances of the  quadratures $\mathcal Q$ (full orange line) and $\mathcal P$ (dashed blue line) versus medium thickness, for an input saturation level $s(0)=1$ in the case of (a) a shot noise limited coherent input state, (b) a 10\,dB $\cal P$--squeezed and (c) a 10\,dB $\cal Q$--squeezed vacuum input states. The grey dot-dashed horizontal line corresponds to the standard quantum limit.  \label{Noise}}
\end{figure}

In both equations, the first term simply represents the propagation of the input squeezing spectrum ${\cal S}_X\left(0,\nu\right)$. The second one is related to the influence of the medium noise correlations $\left\langle F_{\mu\nu}\left(z_1,t_1\right)F_{\alpha\beta}\left(z_2,t_2\right)\right\rangle$. The diffusion coefficients, which can be obtained with the generalized Einstein relation (see Eq.\,\ref{Langevin}), are given in Table\,\ref{Tab}. The $z-$dependence of the variances can then be deduced from the terms of each quadrature squeezing spectrum, using Eqs.\,(\ref{Langevin},\ref{PropaS},\ref{Lambda}): 

\begin{align}
\begin{split}
{\cal S_{P}}\left(z,\nu\right)= & \frac{1}{G\left(z\right)}{\cal S_{P}}\left(0,\nu\right)+1-\frac{1}{G\left(z\right)}\\&+3s\left(z\right)\ln G\left(z\right)\ ,
\label{NoiseP}
\end{split}\\
\begin{split}
{\cal S_{Q}}\left(z,\nu\right)= & G\left(z\right){\cal S_{Q}}\left(0,\nu\right)-1+G\left(z\right)\\&+\frac{\nu^{2}}{\Gamma_{0}^{2}}\ \frac{3\ln G\left(z\right)-\frac{1}{s\left(0\right)}+\frac{1}{s\left(z\right)}}{s\left(z\right)}\ .
\label{NoiseQ}
\end{split}
\end{align}

These equations demonstrate that some noise is added, whatever the input saturation and the spectrum of the probe field. Indeed, the probe field quadratures couple with non-zero noise correlations because the pump field makes the population of the system incoherently cycle from the ground states to the upper state. The small residual population of the excited state can then decay through spontaneous emission, adding some noise to the probe field quadratures. Figure\,\ref{Noise} shows how the noises of both quadratures evolves in the case of a coherent input state or in the case of $\cal P$--squeezed or $\cal Q$--squeezed vacuum input states. In all three cases, the $\cal Q$--component amplification leads to a noise increase. Moreover, the absorption of the $\cal P$--component makes its noise tend to 1 in the thick medium limit $s(z)\rightarrow 0$. A quantum state propagating in a medium under CPO conditions thus cannot be preserved and undergoes a non-unitary transformation, which cannot be compensated.

\section{Conclusion}
\label{sectionV}
In this article, we have investigated the quantum properties of a probe field propagating in an ultra-narrow CPO configuration in a $\Lambda$-system. To this aim, we have treated the probe field quantum mechanically, while keeping a semi-classical approach for the stronger coupling drive field. Moreover, both the quantum average values and the fluctuations of all quantum observables have been derived analytically at first order in probe field Rabi frequency. We have demonstrated that the small number of atoms that are promoted to the upper level of the $\Lambda$-system leads to spontaneous emission, which is sufficient to destroy the quantum noise properties of the input probe field. We have illustrated this feature by considering several squeezed states of light incident on the medium. In all cases, the variances of the quadratures at the output of the medium exceed the standard quantum limit, showing that squeezing is destroyed.

This conclusion contradicts the statement that phase sensitive amplification automatically generates mean that non-classical states of light. In our system, although the net gain depends on which quadrature is detected, a feature that is reminiscent of phase sensitive amplifiers, the quadrature whose power decreases with propagation is not ``de-amplified'', but genuinely absorbed.

Moreover, some years ago the investigation of quantum noise properties under slow and fast light propagation \cite{Boyd10} showed that for an ideal gain medium the noise figure is always less than two and can be set to 1, while a loss medium arbitrarily increases the noise because of the random loss of photons. We demonstrate here that in the case of CPO, a quantum noise degradation always arises because of spontaneous emission, even when the transmission is more than 1. Indeed, since the CPO phenomenon originates from the saturation of absorption along the two legs of the $\Lambda$-system, it is unavoidably accompanied by a small population in the upper level. Although our initial guess was that this population is so small that it can be neglected for quantum storage using ultra-narrow CPO, it appears that it is sufficient to completely spoil the quantum properties of light. This phenomenon should be kept in mind when using resonant atomic systems to create squeezed light, for example via quasi-resonant four-wave mixing. Although the detuning from resonance might be thought to be large enough to make the excited level population negligible, one should pay particular attention to the spontaneous emission induced by such small excitation of the system.
\selectlanguage{english}

\begin{acknowledgments}
The authors acknowledge funding by Indo-French CEFIPRA, Labex PALM,
R\'egion IdF DIM Nano-K, and Institut Universitaire de France.
\end{acknowledgments}

\bibliography{bibliography}
\bibliographystyle{apsrev4-1}

\appendix

\section{Link between envelope operator $\cal E$ and annihilation operator $a$}
\label{appendixA}
This appendix gives some details about the operator $\cal E$ used in Eq.\,(\ref{TotalField}). $\cal{E}$ is the dimensionless complex amplitude of the propagating probe field $\mathbf{E}_p$, written in a frame rotating at $\omega_0$:
\begin{equation}
\mathbf{E}_p\left(z,t\right)=\sqrt{\frac{\hbar\omega_{0}}{2\epsilon_{0}V}}\left({\cal E}\left(z,t\right)e^{-\mathrm{i}\omega_{0}\left(t-\frac{z}{c}\right)}\mathbf{e}_{||}+\mbox{h.c.}\right)\ .
\end{equation}

In the continuous limit ($V\rightarrow\infty$), the relation between this envelope operator and the operators acting on the electromagnetic field is \cite{Loudon}
\begin{align}
{\cal E}\left(z,t\right)=&+\mathrm{i}\sqrt{\frac{L}{2c\pi}}\int_{-\infty}^{+\infty}a\left(\omega\right)e^{-\mathrm{i}\left(\omega-\omega_{0}\right)\left(t-z/c\right)}\mathrm{d}\omega\ , \nonumber\\
{\cal E}\left(z,t\right)^\dagger=&-\mathrm{i}\sqrt{\frac{L}{2c\pi}}\int_{-\infty}^{+\infty}a^\dagger\left(\omega\right)e^{+\mathrm{i}\left(\omega-\omega_{0}\right)\left(t-z/c\right)}\mathrm{d}\omega\ ,\label{eq:E}
\end{align}
where $V=L^3$ is the quantization volume, and the commutation relation for the electromagnetic field operators is $\left[a(\omega_1),a^\dagger(\omega_2)\right]=\delta(\omega_1-\omega_2)$. $\cal E$ and $\cal E^{\dagger}$ are defined in a frame rotating at $\omega_0$, as superpositions of the annihilation and creation operators $a$ and $a^{\dagger}$, respectively. Their spectral components are given by:
\begin{align}
{\cal E}_{p}\left(z,\nu\right) & =+\mathrm{i}\sqrt{\frac{L}{c}}e^{\mathrm{i}\nu z/c}a_{p}\left(\omega_{0}+\nu\right)\ , \nonumber \\
{\cal E}_{p}^{\dagger}\left(z,\nu\right)&=-\mathrm{i}\sqrt{\frac{L}{c}}e^{\mathrm{i}\nu z/c}a_{p}^{\dagger}\left(\omega_{0}-\nu\right)\ .
\label{eq:Enu}
\end{align}
It should be emphasized that ${\cal E}_{p}$ and ${\cal E}_{p}^\dagger$ are not Hermitian conjugate one of the other. Their commutation rules can be deduced from the field operators $a$ and $a^{\dagger}$:
\begin{align}
\left[{\cal E}_p\left(z,\nu\right),{\cal E}_p^\dagger\left(z,\nu'\right) \right]=\frac{L}{c}\delta\left(\nu+\nu'\right)\ .
\label{eq:ECommutation}
\end{align}

The quadratures $\cal P$ and $\cal Q$ of $\cal E$ can then be defined in the Fourier domain by:
\begin{align}
\begin{split}
{\cal P}\left(z,\nu\right)&=\frac{1}{2}\left({\cal E}\left(z,\nu\right)+{\cal E^{\dagger}}\left(z,\nu\right)\right)
\\&
=\mathrm{i}\sqrt{\frac{L}{4c}}e^{\mathrm{i}\nu z/c}\left(a\left(\omega_0+\nu\right)-a^\dagger\left(\omega_0-\nu\right)\right),
\end{split} \nonumber \\
\begin{split}
{\cal Q}\left(z,\nu\right)&=\frac{1}{2\mathrm{i}}\left({\cal E}\left(z,\nu\right)-{\cal E^{\dagger}}\left(z,\nu\right)\right)\\&
=\mathrm{i}\sqrt{\frac{L}{4c}}e^{\mathrm{i}\nu z/c}\left(a\left(\omega_0+\nu\right)+a^\dagger\left(\omega_0-\nu\right)\right).
\end{split}
\end{align}
$\cal P$ and $\cal Q$ can thus have a complex amplitude.

Using the commutation relation (\ref{eq:ECommutation}), it is possible to compute the squeezing spectrum for vacuum:
\[
\begin{array}{l}
\int\mathrm{d}\nu'\left\langle {\cal P}\left(z,\nu\right){\cal P}\left(z,\nu'\right)\right\rangle
\\
\,\,\,\,\,\,= \frac{1}{4}\int\mathrm{d}\nu'\left\langle {\cal E}^{\dagger}\left(z,\nu\right){\cal E}\left(z,\nu'\right)+{\cal E}\left(z,\nu\right){\cal E}^{\dagger}\left(z,\nu'\right)\right\rangle
\\
\,\,\,\,\,\,= \left(\frac{L}{4c}+\frac{1}{2}\int\mathrm{d}\nu'\left\langle {\cal E}^{\dagger}\left(z,\nu\right){\cal E}\left(z,\nu'\right)\right\rangle \right)
\\
\,\,\,\,\,\,= \frac{L}{4c}.
\end{array}
\]
so that the Eq.\,(\ref{eq:Flux}) leads to ${\cal S}_{X}\left(\omega\right)=1$.

\section{Transmission coefficient definitions}
\label{appendixB}
The transmission coefficients $T_{\Theta=0,\frac{\pi}{2}}$ are considered after propagation in the whole medium. They are thus related to the $4\times4$ transfer matrix per unit length $\mathcal{T}\left(z\right)$ given in Eq. $\left(10\right)$ of Ref.\cite{Neveu17} by
\begin{align*}
T_{\Theta=0} & =\exp\left[\int_{0}^{L}dz\;\mathcal{T}_{44}\left(z\right)\right]\ ,\\
T_{\Theta=\frac{\pi}{2}} & =\exp\left[\int_{0}^{L}dz\;\mathcal{T}_{33}\left(z\right)\right]\ .
\end{align*}
\end{document}